\documentclass[11pt,twoside]{article}

\usepackage{asp2004}
\usepackage{epsf}
\usepackage{epsfig}
\usepackage{lscape}
\usepackage{amsmath}
\usepackage{amssymb}

\begin{document}
\title{Photometric Calibration of the Supernova Legacy Survey Fields}
\author{Nicolas Regnault,\\
for the SNLS Collaboration}
\affil{LPNHE - Laboratoire de Physique Nucl\'eaire et de Hautes-Energies\\
IN2P3 - CNRS - Universit\'es Paris VI et Paris VII\\
4 place Jussieu\\
Tour 33 - Rez de chauss\'ee\\
75252 Paris Cedex 05}

\begin{abstract} The 5-year project Supernova Legacy Survey (SNLS) delivers $\sim 100$
 Type-Ia supernovae (SNe~Ia) per year, in the redshift range $0.3 < z
 < 1.0$, with well-sampled $g'r'i'z'$ lightcurves. The SNLS
 Collaboration uses the 1 deg$^2$ Megacam imager (36 $2048 \times
 4612$ thinned CCDs) mounted on the 3.6-m Canada-France-Hawaii
 Telescope (CFHT) to observe four fields around the sky, in four
 filters. The primary goal of the project is to measure the dark
 energy equation of state with a final statistical precision of $\pm
 0.05$. We have shown, using the first year dataset that the
 calibration uncertainties are currently the dominant contribution to
 the systematic error budget.
 
 The calibration of the SNLS dataset is challenging in several
 aspects. First, Megacam is a wide-field imager, and only a handful of
 its 36 CCDs can be directly calibrated using standard star
 observations. Second, measuring the rest-frame $B$-band luminosity of
 SNe~Ia over the $0.3<z<1.0$ redshift range requires an excellent flux
 intercalibration of the Megacam bands. Finally, the SN~Ia SED differs
 significantly from that of stars and transfering the stellar
 calibration to the SNLS data requires a precise knowledge of the
 SN~Ia spectra and the instrument transmissions.
 
 We present and discuss the SNLS calibration strategy used to analyze
 the first year data set. We present the calibration aspects which
 impact most the cosmological measurements. We also discuss the
 intercalibration of the SNLS with other surveys, such as the
 CFHTLS-Wide and the SDSS.
\end{abstract}

\section{The Supernova Legacy Survey}

 {Type Ia supernovae} (SNe~Ia) are a powerful probe of the history of
 cosmic expansion. The first distant SN~Ia surveys
 \citep{Perlmutter97, Perlmutter99, Riess98b} detected the acceleration
 of the expansion, and provided strong evidence for repulsive {dark
 energy} driving the expansion. Subsequent surveys \citep{Knop03,
 Tonry03, Barris04, Riess04} confirmed this result. Determining
 the nature of dark energy by measuring its equation of state,
 i.e. its pressure over density ratio: $w = p / \rho$ has now become a
 central question in observational cosmology. Many dark energy models
 have been proposed, besides the historical cosmological constant
 ($w=-1$). Some of them predict values of $w$ significantly different
 from $-1$. Unfortunately, the best constrains obtained from the
 SNe~Ia surveys mentioned above are consistent with a wide range of
 dark energy models.
 
 Improving them to the point where $w=-1$ could be excluded or
 confirmed requires a ten-fold larger sample, i.e. O(1000) SNe at $0.3
 < z < 1.0$ ---where $w$ is best measured--- in order to improve not
 only on statistics, but also on systematics. Several
 second-generation surveys have been designed to build such samples:
 the {Supernova Legacy Survey} (SNLS), at the Canada-France-Hawaii
 Telescope (CFHT), and the ESSENCE project at the Cerro-Tololo
 InterAmerican Observatory. 
 
 The Supernova Legacy Survey delivers $\sim 100$ SNe~Ia per
 year, with well sampled $g'r'i'z'$ lightcurves. Over the five year
 duration of the project, we expect to obtain several hundred SNe~Ia,
 all spectroscopically identified. The SNLS project is comprised of
 two components: a large imaging survey to detect supernovae and
 monitor their lightcurves, and a spectroscopic survey, to confirm the
 nature of the candidates and measure their redshift.
 
 The imaging survey is a component of the larger CFHT Le\-ga\-cy
 Sur\-vey pro\-ject~\citep{cfhtls}. The CFHTLS operates the one
 square-degree imager MEGACAM~\citep{MegacamPaper} mounted on the
 prime focus of the Canada France Hawaii Telescope, and has been
 allocated 474 nights over 5 years. The whole project actually
 consists of 3 distinct surveys: a very wide shallow survey (1300
 square degrees), a wide survey (120 square degrees) and a deep survey
 (4 square degrees). The 4 pointings of the deep survey
 (Table~\ref{table:deep_fields}) are evenly distributed in right
 ascension, and observed at five equally space epochs during a Megacam
 Run, which lasts at least 14 nights around new moon. The observations
 are taken in a combination of the $r$, $i$ plus $g$ and
 $z$ megacam filters, depending of the phase of the moon.

\begin{table}
\centering
\begin{tabular}{ccc|c}
 Field &  RA(2000)    & Dec(2000)  &  $E_{B-V}$ (MW)    \\
\noalign{\smallskip}
\tableline
\noalign{\smallskip}
D1     &  02:26:00.00 & $-04$:30:00.0 & 0.027\\
D2     &  10:00:28.60 & +02:12:21.0 & 0.018\\
D3     &  14:19:28.01 & +52:40:41.0 & 0.010\\
D4     &  22:15:31.67 & $-17$:44:05.0 & 0.027\\
\noalign{\smallskip}
\tableline
\end{tabular}
\caption{Coordinates and average Milky Way extinction  \citep[from][]{Schlegel98}  of 
fields observed by the Deep/SN component of the CFHTLS.\label{table:deep_fields}       
}
\end{table}

 From the first year of operations, we obtained 71 type Ia supernovae
 spectroscopically identified and well sampled enough to be placed on
 a Hubble diagram. Using this unique data set, supplemented with 41
 published low-redshift SNe~Ia, we have built a {Hubble diagram}
 extending to $z=1$, with all distance measurements involving at least
 two bands \citep{astier06}. The cosmological fit to this first year
 SNLS Hubble diagram gives the following results : $\Omega_m = 0.263 \pm
 0.042\;(stat) \pm 0.032\;(sys)$ for a flat $\Lambda$CDM model; and $w
 = -1.023 \pm 0.090\;(stat) \pm 0.054\;(sys)$ for a flat cosmology
 with constant equation of state $w$, when combined with the
 constraint from the recent {Sloan Digital Sky Survey} measurement of
 baryon acoustic oscillations. This is currently the best available
 constraint on the dark energy equation of state.
 
 Improving significantly this result requires to push down the
 systematic uncertainties. In \citet{astier06} we have shown that
 photometric calibration is currently the dominant source on the cosmological
 parameter error budget. In the following of this paper we discuss our
 current calibration strategy, and our efforts to improve the
 calibration of the survey. In section \ref{sec:generalities}, we
 present the photometric calibration constraints of a supernova survey. We then
 describe the Megacam imager (section
 \ref{sec:megacam}). We describe the calibration of the
 first year data (section \ref{sec:calibration_procedure}). Finally,
 we discuss our current efforts to calibrate the SNLS survey using
 Sloan and HST secondary and primary standards, which should allow us
 to cross-check the {Vega/Landolt zero-points}, and more accurately
 calibrate $z$-band observations.

\section{Calibrating a Dark-Energy Survey\label{sec:generalities}}
 
 Currently broadband photometric measurements are calibrated using
 observations of reference standard stars, which define a photometric
 system \citep[see][]{Landolt92, Smith02}. Most photometric systems
 are ultimately tied to the SED of {Vega}. However, since Vega is
 $10^6$ times brighter than the mag 15 secondary
 standards currently used by large telescopes, the path from the Vega
 SED determination to the zero points of modern standard star catalogs
 is rather indirect. This can lead to systematic errors of a few
 percent when trying to convert magnitudes into fluxes
 \citep{Fukugita96}.
 
 In supernova cosmology, we study the
 luminosity-distance-versus-red\-shift-re\-lation, $d_L(z)$. In order
 to measure the luminosity distance of a supernova, we have to {\em
 infer} its apparent peak brightness at some chosen reference
 wavelength in the supernova {\em rest-frame}, using photometric
 measurements performed at a few fixed passbands in the observer
 frame. Such a transformation is called a k-correction.  In order to
 k-correct supernova magnitudes, we need the following ingredients:

\begin{enumerate}
 \item the zero points of the photometric system used to calibrate our
 measurements, in order to convert calibrated magnitudes into
 fluxes. Depending on how the magnitude system was tied to the SED of
 the fundamental flux standard (such as Vega), this step can be one of
 the dominant sources of systematic errors on the cosmological
 measurements. Note however that only the band-to-band relative values
 of the zero points have an impact on cosmology (i.e. the Vega colors
 in the photometric system used to calibrate the measurements).

 \item a model of the instrument passbands including the transmission
 of the optics, the mirror reflectivity, the filter transmissions, the
 CCD quantum efficiency and finally, a model of the atmospheric
 absorption, especially in the near infrared, where atomspheric
 absorption lines are rather strong and in the near-UV.

 \item a model of the effective passbands of the photometric system
 used to calibrate the survey. These passbands usually differ from
 those which equip the survey telescope.
 
 \item a model of the supernova SED as a function of time \citep[see
 for example][]{Guy05}. We won't address this issue here.

\end{enumerate}

 \paragraph{} The {supernovae} discovered by the SNLS cover the $0.3 < z
 < 1.0$ redshift range. This dataset must be supplemented by an
 additional set of SNe~Ia at much lower redshift ($z \sim 0.05$), in
 order to extract precise measurements of the cosmological parameters
 from a Hubble diagram. Most well studied nearby SNe~Ia were
 discovered by \citep{Hamuy96,Riess99} and others during the last
 decade, and calibrated in the photometric system defined by
 \citep{Landolt92}.  We therefore have to adopt the same calibration
 source for the SNLS sample. This avoids introducing additional
 systematic uncertainties between the distant and nearby SN
 fluxes. However, this is complicated by the fact that the Megacam
 passbands differ significantly from the {$UBVRI$} filters used by
 Landolt.

\section{The Megacam Imager\label{sec:megacam}}

 Megacam is a wide-field imager, built by the Commissariat \`a
 l'\'Energie Atomique (CEA) for the prime focus of the 3.6-m Canada
 France Hawaii telescope. It covers a field area of $0.96 \times 0.94$
 deg$^2$, with an excellent and remarkably uniform image quality. The
 focal plane is made of 36 thinned $2048\times4612$ CCDs, with pixels
 of 13.5 $\mu m$ that subtend 0.18 arcsec on a side. The whole focal
 plane comprises $\sim 340$ million pixels. However, it is read in
 less than 40 seconds. Each CCD is read out from two amplifiers.

 Great care has been taken in the internal calibration of the
 imager. Indeed, any not accounted-for shutter imperfection or
 non-linearity of the detector/electronics can bias the measurements
 and hence the cosmology. Another potential source of problems is the
 uniformity of the camera: we cannot afford to calibrate each of the
 36 CCDs using standard star observations. Most standard stars are
 therefore observed with the center CCDs, and the camera
 non-uniformities of the photometric response are carefully
 modeled. Any residual radial non-uniformities of the photometric
 response may distort the supernova luminosity distribution, and bias
 the cosmological measurements. In this section, we review the
 critical camera systems. In the next section, we will present the photometric 
 calibration procedure.

 \paragraph{\rm \em The Shutter system} The {shutter precision} is a
 potential source of systematic uncertainties, given (1) the possible
 non uniformities due to the shutter motion and (2) the exposure time
 differences between the calibration exposures (a few seconds) and the
 science exposures (hundreds of seconds). The shutter system was
 carefully designed in order to ensure that (1) the accuracy of the
 exposure time measurement is better than 1 ms and (2) the uniformity
 of the exposure time is to better than 1\%. The design of the shutter
 is based on the controlled rotation of a half disk, --- one meter in
 diameter --- in order to ensure a constant speed when the shutter
 crosses the CCD mosaic. The exact duration of the exposure time is
 measured with a precision of 0.5 ms with a dedicated system
 independent from the shutter motion controller. The shutter precision
 was investigated by the CFHT team. It was shown that the
 non-uniformity due to the shutter is below 0.3\% accross the
 mosaic. The systematic flux differences between the exposures were
 found to be below 1\% (r.m.s.).
 
 \paragraph{\rm \em Linearity} The {linearity of the CCDs} and the
 readout electronics is also a potential source of systematic
 uncertainties. The requirements stated that the linearity of each
 channel had to be better than 1\%. The linearity was later
 investigated by the CFHT team. It was found to be within the
 specifications, except for CCD\#17.
 
 \paragraph{\rm \em Filters} The filter system is a juke box which
 holds up to 8 filters. The CFHT Legacy Survey performs obervations in
 five bands, labeled {$u_M, g_M, r_M, i_M, z_M$}, similar to the
 SDSS {$u'g'r'i'z'$} bands. The SNLS uses only the $g_M,r_M,i_M$- and
 $z_M$-band observations. The filters currently mounted on Megacam
 are interferometric filters manufactured by REOSC/Sagem. Their
 transmissions were characterized by the manufacturer and the CFHT
 team. Small systematic differences were found between the Megacam and
 SDSS filters, which translate into small color terms between both
 instruments (see below).

\section{The Photometric Calibration Procedure\label{sec:calibration_procedure}}
 
 \subsection{Elixir Pipeline. Uniformity of the Photometric Response} 

 At the end of each CFHT run, the raw images are processed using the
 Elixir pipeline, developed by the CFHT team \citep{Magnier04}. Master
 flat field images and fringe corrections are built from all the data
 taken during the run, including PI data. The Elixir pipeline applies
 these flat fields to the data, subtract the fringe patterns and
 determines an astrometric solution.

\begin{figure}[!t]
 \centering
 \includegraphics[width=0.47\linewidth]{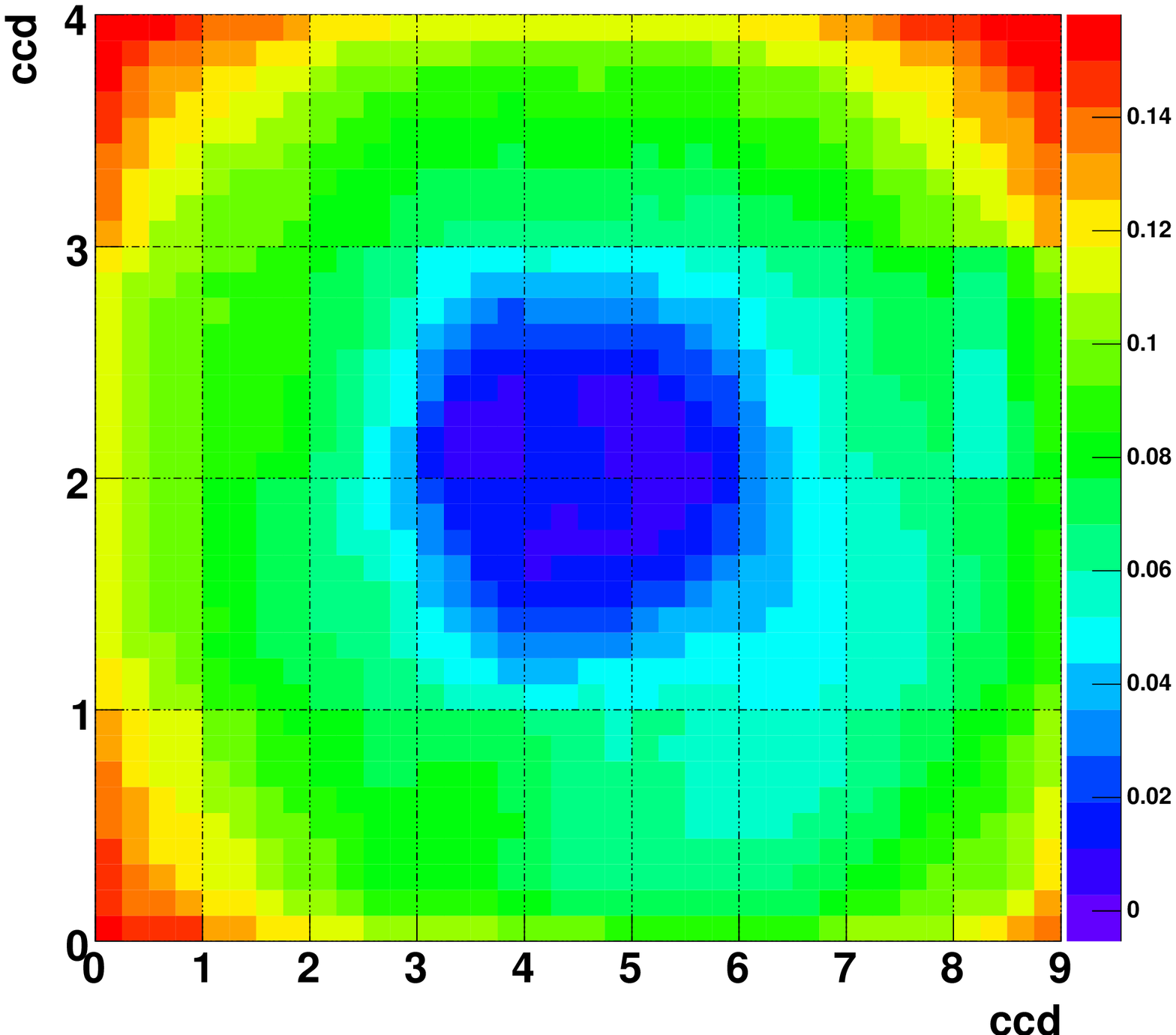}
 \includegraphics[width=0.47\linewidth]{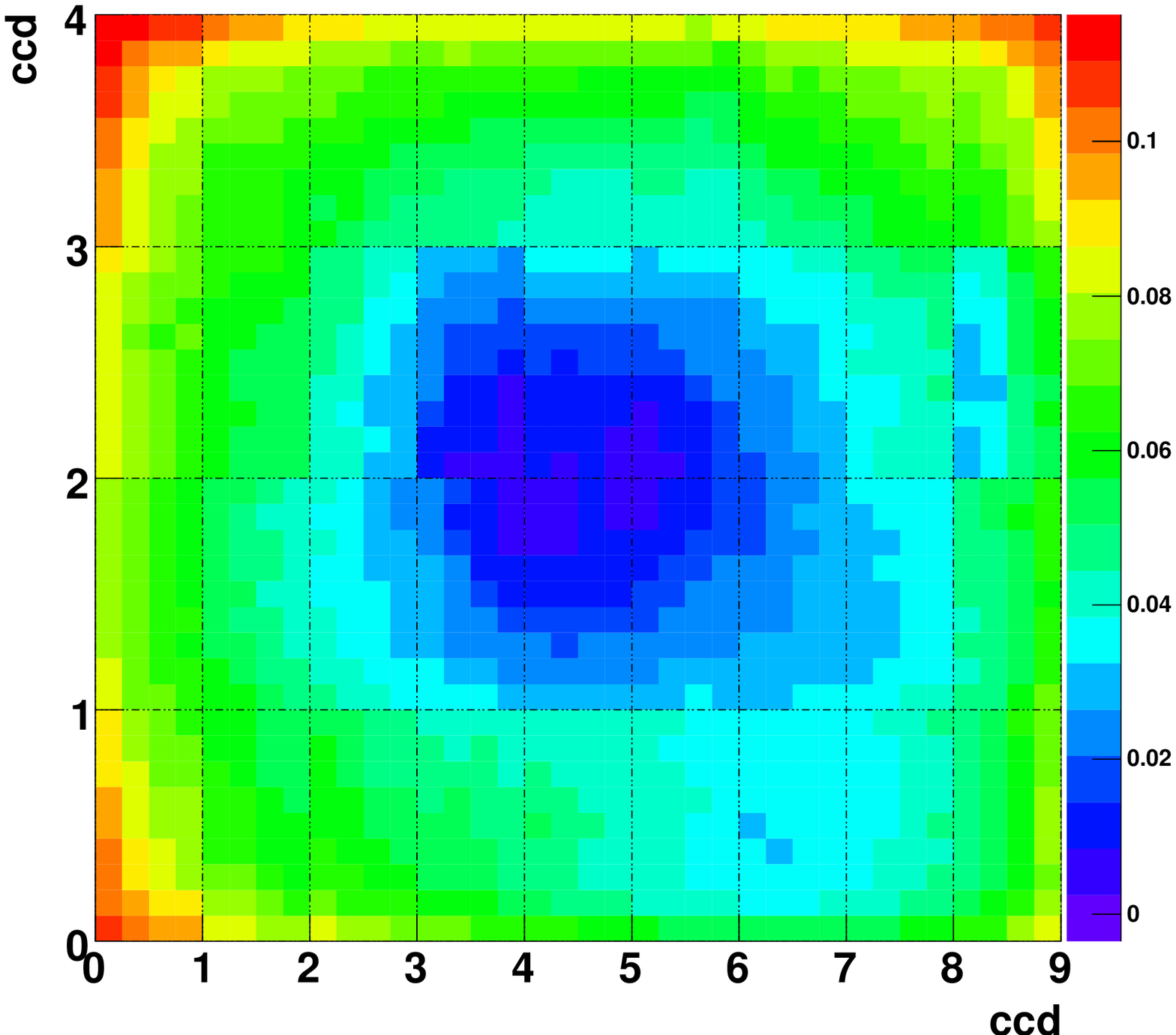}
 \caption{Map of the non-uniformities of the photometric response, in
 the $r$-band (left) and $i$-band (right). Each map represents the
 whole focal plane. The grids materialize the CCDs. The maps are
 determined using dithered exposures of a dense stellar field. Each
 CCD is divided into $4 \times 9$ cells. Since each star of the dense
 stellar field is observed on several cells during the dithering
 sequence, we are able to intercalibrate each cell with respect to a
 cell chose as a
 reference. \label{fig:photometric_non_uniformities}}
\end{figure}

 {Flat-fielding} ensures that the pixel-to-pixel response is uniform
 accross the entire focal plane. However, it was found that the
 photometric response measured on flat-fielded images was not
 uniform. In other words, two measurements of the same star, at two
 different locations of the focal plane, may yield different
 instrumental fluxes. These non-uniformities have been measured using
 dithered observations of dense stellar fields, and have been found to
 be radial and as large as 15\% (Fig.~\ref{fig:photometric_non_uniformities}). These observed
 non-uniformities may have a number of explanations, and a combination
 of explanations is likely. One is the geometric distorsion: a pixel
 at the center of the field subtends a different solid angle on the sky than a
 pixel located on the edges. The flat field provides a uniform
 illumination, and does not account for this effect. However, such an
 effect would be achromatic, which is not the case here. Moreover, the
 amplitude of the geometric distorsion, which can be determined from
 the astrometry only accounts for half of the observed effect. Another
 possible explanation is scattered light during the flat field
 observations.

 Modeling scattered light, or removing it totally is extremely
 difficult. Therefore the CFHT team has chosen to measure the
 non-uniformities of the photometric response using the dense stellar
 field dithers mentioned above, and to include this model into the
 flat field corrections. The Elixir team has deliberately chosen to
 provide reduced data which has a uniform photometric response across
 the mosaic, at the expense of a non-uniform sky background.

 In order to improve significantly the measurements published in
 \citep{astier06}, SNLS has started an ambitious program to decrease
 the internal calibration uncertainties down to 1\%.  A considerable
 amount of work is therefore still being carried out on this subject
 by the Elixir and SNLS teams. In particular, the dense stellar field
 dithers are being reanalyzed by both teams using different methods,
 in order to improve the photometric correction function, and to
 investigate potential non-uniformities in the filter passbands.

 \subsection{Building Tertiary Standard Catalogs}
 
 The images preprocessed by the Elixir pipeline have a flat
 photometric response. Each image has then to be aligned on the
 Landolt catalog. We have chosen to proceed in two steps: first, we
 have built a catalog of so called tertiary standards, i.e. science
 field stars whose fluxes have been calibrated using Landolt star
 observations. Then, using this catalog, each image containing
 tertiary standards can be calibrated.

 Building a tertiary standard catalog is relatively easy since we have
 between 12 and 25 epochs (depending on the passband: 12 in $g_M$ and
 $z_M$ and over 20 in $r_M$ and $i_M$) for each science field, and
 since both standard and science fields were repeatedly observed.
 Photometric nights were selected using
 the CFHT ``Skyprobe'' instrument \citep{SkyProbe}, which monitors
 atmospheric transparency in the direction that the telescope is
 pointing. Only the 50\% of nights with the smallest scatter in
 transparency were considered. For each night, stars were selected in
 the science fields and their aperture fluxes measured and corrected
 to an airmass of 1 using the average atmospheric extinction of Mauna
 Kea.  These aperture fluxes were then averaged, allowing for
 photometric ratios between exposures of the same night.  Stable
 observing conditions were indicated by a very small scatter in these
 photometric ratios (typically 0.2\%); again the averaging was robust,
 with 5-$\sigma$ deviations rejected.  Observations of the {Landolt
 standard star fields} were processed in the same manner, though their
 fluxes were not averaged.  The apertures were chosen sufficiently
 large (about 6\arcsec\ in diameter) to bring the variations of
 aperture corrections across the mosaic below 0.005 mag. However,
 since fluxes are measured in the same way and in the same apertures
 in science images and standard star fields, we did not apply any
 aperture correction.

 Using standard star observations, we first determined zero-points by
 fitting linear color transformations and zero-points to each night
 and filter, however with color slopes common to all nights.  In order
 to account for possible non-linearities in the Landolt to MegaCam
 color relations, the observed color-color relations were then
 compared to synthetic ones derived from spectrophotometric
 standards. This led to shifts of roughly 0.01 in all bands other than
 $g_M$, for which the shift was 0.03 due to the nontrivial relation
 to $B$ and $V$.

\begin{figure}[!t]
 \centering
 \includegraphics[width=0.8\linewidth]{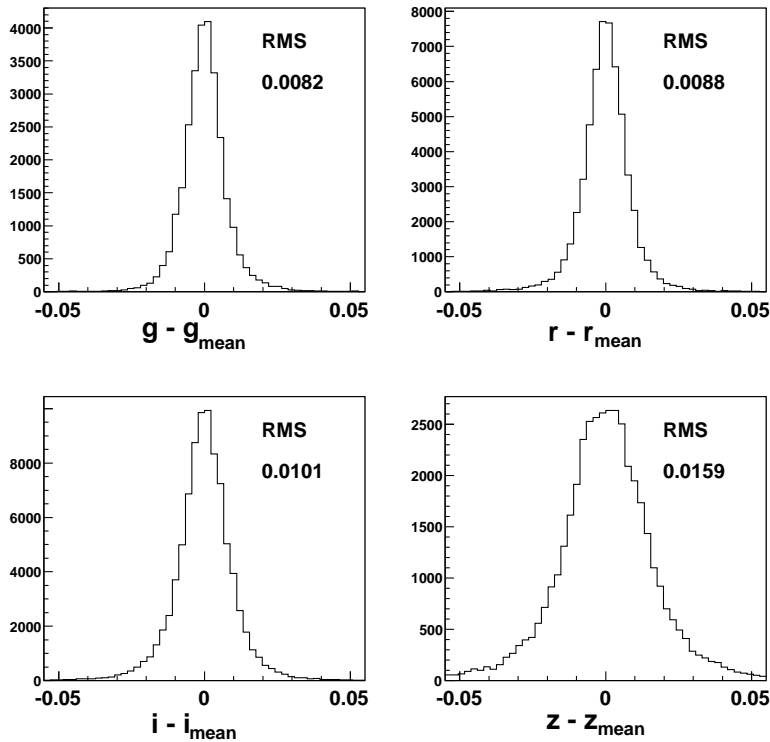}
 \caption{The calibration residuals --- {\em i.e.} the residuals around
 the mean
 magnitude of each Deep field tertiary
 standard--- in the bands $g_M$, $r_M$, $i_M$ and $z_M$,
 for all CCDs and fields, with one entry per star and epoch.
 The dispersion is below 1\% in $g_M$, $r_M$ and $i_M$,
 and about 1.5\% in $z_M$.
 \label{figure:calibration_residuals}}
 \end{figure}

 We then applied the zero-points appropriate for each night to the
 catalog of science field stars of that same night. These magnitudes
 were averaged robustly, rejecting 5-$\sigma$ outliers, and the
 average standard star observations were merged.  Figure~\ref{figure:calibration_residuals} shows the dispersion of the
 calibration residuals in the $g_M$, $r_M$, $i_M$ and $z_M$ bands.
 The observed standard deviation, which sets the upper bound to the
 repeatability of the photometric measurements, is about or below
 0.01~mag in $g_M$, $r_M$ and $i_M$, and about 0.016~mag in $z_M$.

 For each of the four SNLS fields, a catalog of tertiary standards was
 produced using the procedure described above. The dominant
 uncertainty in the photometric scale of these catalogs comes from the
 determination of the color-color relations of the standard star
 measurements. For the $g_M$, $r_M$ and $i_M$ bands, a zero-point
 offset of 0.01 mag would easily be detected; hence we took this value
 as a conservative uncertainty estimate. The $z_M$ band is affected
 by a larger measurement noise, and it is calibrated with respect to
 $I$ and $R-I$ Landolt measurements.  We therefore attributed to it a
 larger zero point uncertainty of 0.03 mag.

 Once magnitudes are assigned to tertiairy standards, supernova
 magnitudes are measured by estimating the supernova flux and the
 field stars (i.e. the tertiary standard) fluxes with the same PSF
 photometry.  Although supernovae involve a differential photometry
 and field stars do not, we were able to prove that the possible
 biases of the supernova to field stars flux ratios are negligible
 \citep[see][]{astier06}
 
 \paragraph{\rm \em Megacam Filter Model} For the MegaCam filters, we
 used the measurements provided by the manufacturer, multiplied by the
 CCD quantum efficiency, the MegaPrime wide-field corrector
 transmission function, the CFHT primary mirror reflectivity, and the
 average atmospheric transmission at Mauna Kea. As an additional
 check, we computed synthetic MegaCam-SDSS color terms using the
 synthetic transmissions of the SDSS 2.5-m telescope
 \citep{SkyServerInstrument} and {spectrophotometric standards} taken
 from \citep{Pickles98, Gunn83}. Since the SDSS science catalog
 \citep{Finkbeiner94,Raddick02,SkyServer} shares thousands of objects
 with two of the four fields repeatedly observed with MegaCam, we were
 able to compare these synthetic color transformations with the
 observed transformations. We found a good agreement, with
 uncertainties at the 1\% level.  This constrains the central
 wavelengths of the MegaCam band passes to within 10 to 15~\AA\ with
 respect to the SDSS 2.5m band passes.

 \paragraph{\rm \em Landolt Effective Filter Model} The choi\-ce of
 filter band pas\-ses to use for Landolt based observations is not
 unique.  Most previous supernova cosmology works assumed that the
 determinations of \citep{Bessel90} describe the effective Landolt
 system well, although the author himself questions this fact,
 explicitly warning that the Landolt system {\em ``is not a good match
 to the standard system''} -- {\em i.e.} the historical
 {Johnsons-Cousins system}.  Fortunately, \citep{Hamuy92, Hamuy94}
 provide spectrophotometric measurements of a few objects measured in
 \citep{Landolt92}; this enabled us to compare synthetic magnitudes
 computed using Bessell transmissions with Landolt measurements of the
 same objects. This comparison reveals small residual color terms
 which vanish if the $B$, $V$, $R$ and $I$ Bessell filters are
 blue-shifted by 41, 27, 21 and 25~\AA\ respectively. Furthermore, if
 one were to assume that the Bessell filters describe the Landolt
 system, this would lead to synthetic MegaCam-Landolt color terms
 significantly different from the measured ones; the blue shifts
 determined above bring them into excellent agreement. We therefore
 assumed that the Landolt catalog magnitudes refer to blue-shifted
 Bessell filters, with a typical central wavelength uncertainty of 10
 to 15~\AA, corresponding roughly to a 0.01 accuracy for the color
 terms.

\begin{figure}[!t] 
 \centering
 \includegraphics[width=0.9\linewidth]{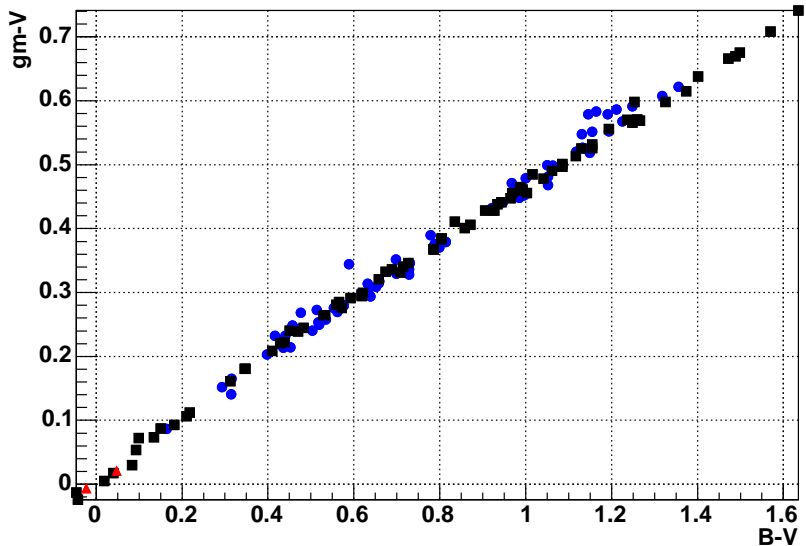}
 \caption{$g_M-V$ vs. $B-V$ color-color plots, built from (1) Megacam
 observations of Landolt stars (blue dots) (2) synthetic $g_M, B,
 V$-band magnitudes computed using our passband models and spectra
 taken from \citep{Gunn83} (red dots) and \citep{Pickles98} (black
 dots). \label{fig:color_color_plots}} \end{figure}

 A powerful check of (1) our alignement on the Landolt system (2) our
 model of the Megacam passbands and (3) our model of the effective
 Landolt filters is to compare observed and synthetic color-color
 plots, using our observations of the Landolt stars and synthetic
 Landolt and Megacam magnitudes of stellar spectra. Figure
 \ref{fig:color_color_plots} presents the $g_M - V$ vs. $B-V$ diagram
 built from (1) Megacam observations of Landolt stars (blue dots) (2)
 synthetic $g_M, B, V$-band magnitudes computed using our passband
 models and spectra taken from \citep{Gunn83} (red dots) and
 \citep{Pickles98} (black dots). We notice an excellent agreement
 between the synthetic and observed magnitudes.
 
\section{Improving the Calibration Path\label{sec:improving_calibration_path}}
 
 As discussed in section \ref{sec:generalities}, the current
 calibration program is not optimal for supernova cosmology: we know
 little about the systematic uncertainties associated with converting
 Landolt magnitudes into fluxes, and there is no official
 determination of the Landolt passbands. Finally, the $z_M$-band is
 redder than the reddest Landolt band ($I$). Therefore, the Megacam to
 Landolt $z$ to $I$ transformation is an extrapolation, based on the
 sampling of the main sequence stars published by
 \citep{Pickles98, Gunn83}.
 
 The calibration landscape is rapidly evolving, and quickly getting
 much richer and redundant. The set of HST flux standards, aligned on
 the white dwarf flux scale \citep{Bohlin04,Calspec} is constantly
 expanding, and includes fundamental calibrators such as {Vega}
 \citep{Bohlin04b} and {{BD +17 4708}} (AB fundamental standard)
 \citep{Bohlin04c}. The SDSS 2.5-m system which has become a de-facto
 standard photometric system is slowly being tied to the white dwarf
 flux scale.

 The SNLS collaboration has therefore started a dedicated effort to
 produce a definitive set of tertiary standards, calibrated against
 these new sets of standards.  This effort consist in observing in all
 bands a dithered sequence of the SNLS fields, parts of the well
 calibrated SDSS Southern Strip, Landolt calibrators and HST
 fundamental standards. The dither sequence will allow us to check the
 uniformity of the photometric response, and detect possible
 variations of the amplifier gains at the sub-percent level. The
 combination of celestial calibrators will permit to calibrate the
 tertiary standards against several important magnitude systems, and
 to check for systematic differences between those systems.
 
\section{Conclusion}

 The calibration of the SNLS
 dataset is challenging in several aspects. First, Megacam is a wide
 field imager, and controlling the uniformity of such an instrument is
 not possible without a dedicated calibration program. We have shown
 in section \ref{sec:calibration_procedure} that not accounted-for
 non-uniformities can bias the cosmological measurements. Another
 difficult task is to control how the Landolt system is tied to Vega,
 its fundamental flux standard. We have therefore embarked in two
 distinct programs to (1) control the internal calibration of our
 imager with a precision better than 1\% and (2) check our current
 calibration path. A longer term goal is to define an absolute flux
 calibration of the SNLS survey, based on the white dwarf flux scale.
 The photometric calibration work currently carried out will allow us 
 to reach the 1\% precision which must be attained to significantly improve the
 measurements of the cosmological parameters.

 \acknowledgements We gratefully acknowledge the assistance of the
CFHT Queued Service Observing Team, led by P. Martin (CFHT). We are
gateful to Jean-Charles Cuillandre for continuous improvement of the
instrument performance. We heavily rely on the team maintaining,
running and monitoring the real-time Elixir pipeline, J-C.~Cuillandre,
E. Magnier and K.~Withington.

\clearpage \thispagestyle{empty} 
\mbox{}
\pagebreak

\end{document}